\documentclass[9pt,twocolumn,twoside]{osajnl}

\journal{ol}

\setboolean{shortarticle}{true} 

\title{Anomalous refraction phenomena in a transparent slab}

\author[1,*]{Michel Lequime}
\author[1]{Claude Amra}

\affil[1]{Aix Marseille Univ, CNRS, Centrale Marseille, Institut Fresnel, Marseille, France}

\affil[*]{Corresponding author: michel.lequime@fresnel.fr}

\ociscodes{(260.2110) Electromagnetic optics; (260.3160) Interference; (120.5710) Refraction; (310.0310) Thin films.}

\begin{abstract}
An exact formulation of the propagation of a monochromatic wave packet impinging upon a transparent, homogeneous, isotropic and parallel slab at oblique incidence is presented. Approximate formulas are derived for low divergence Gaussian light beams. These formulas show the presence of anomalous refraction phenomena at any slab thickness, including negative refraction and flat lensing effects, induced by the reflection on the rear face.
\end{abstract}

\setboolean{displaycopyright}{true}

\begin{document}

\maketitle

\section{Introduction}

Since the first half of the 17$^{\text{th}}$ century and the works of Snell and Descartes on the refraction of a light ray through a boundary between two different isotropic media, the mathematical formula describing the direction change of this light ray when crossing the boundary, the famous sine law, is well established. However, recently, anomalous refraction phenomena were clearly demonstrated, e.g. in a metamaterial made of cascaded metal-dielectric fishnet structures \cite{Valentine2008Nature} or on a nano-structured silicon surface including thin arrays of metallic antennas with a linear phase variation along the interface \cite{Yu2011Science}.

From 2007, in the framework of the development of double-negative metamaterials for photonics \cite{DollingThesis2007,Dolling2007OptExpr}, Dolling emphasized that thin, isotropic and homogeneous silver films, typically a few tens of nanometers thick, had anomalous refraction behavior (negative beam displacements for TM polarization), and that transparent dielectric films with similar thicknesses can exhibit the same type of feature (this time for TE polarization). Moreover, he concluded that hardly any negative refraction occurs for films with increased thickness and assumed that the interference effects of the interfaces become negligible when the thickness of the film is much larger than the wavelength.

In this letter, we show that these interference effects in fact remain present for any film thickness and impact the beam position on the rear face even in the case of bulk material. In Section \ref{sec:Model}, we will describe the main features of our theoretical approach and establish the analytical formula that must be used in this case instead of the classical sine law. Section \ref{sec:Consequences} highlights some important consequences of this formula. Section \ref{sec:Discussion} analyzes the practical conditions to fulfill for allowing experimental demonstration of this anomalous refractive behavior.

\section{Model}
\label{sec:Model}

\subsection{General approach}
\label{sec:GeneralApproach}

Consider first a plane boundary between a semi-infinite incident medium (refractive index $n_0$) and a semi-infinite dielectric medium (refractive index $n_1$), both perfectly transparent. The sine law, i.e.
\begin{equation}
n_0\sin\theta_0=n_1\sin\theta_1
\end{equation}
expresses the conservation of the tangential component of the wave vector at the crossing of the boundary. As a consequence, this relation is only defined for plane waves and its experimental verification must only be based on angular measurement.

\begin{figure}[htbp]
\centering
\includegraphics[width=.7\linewidth]{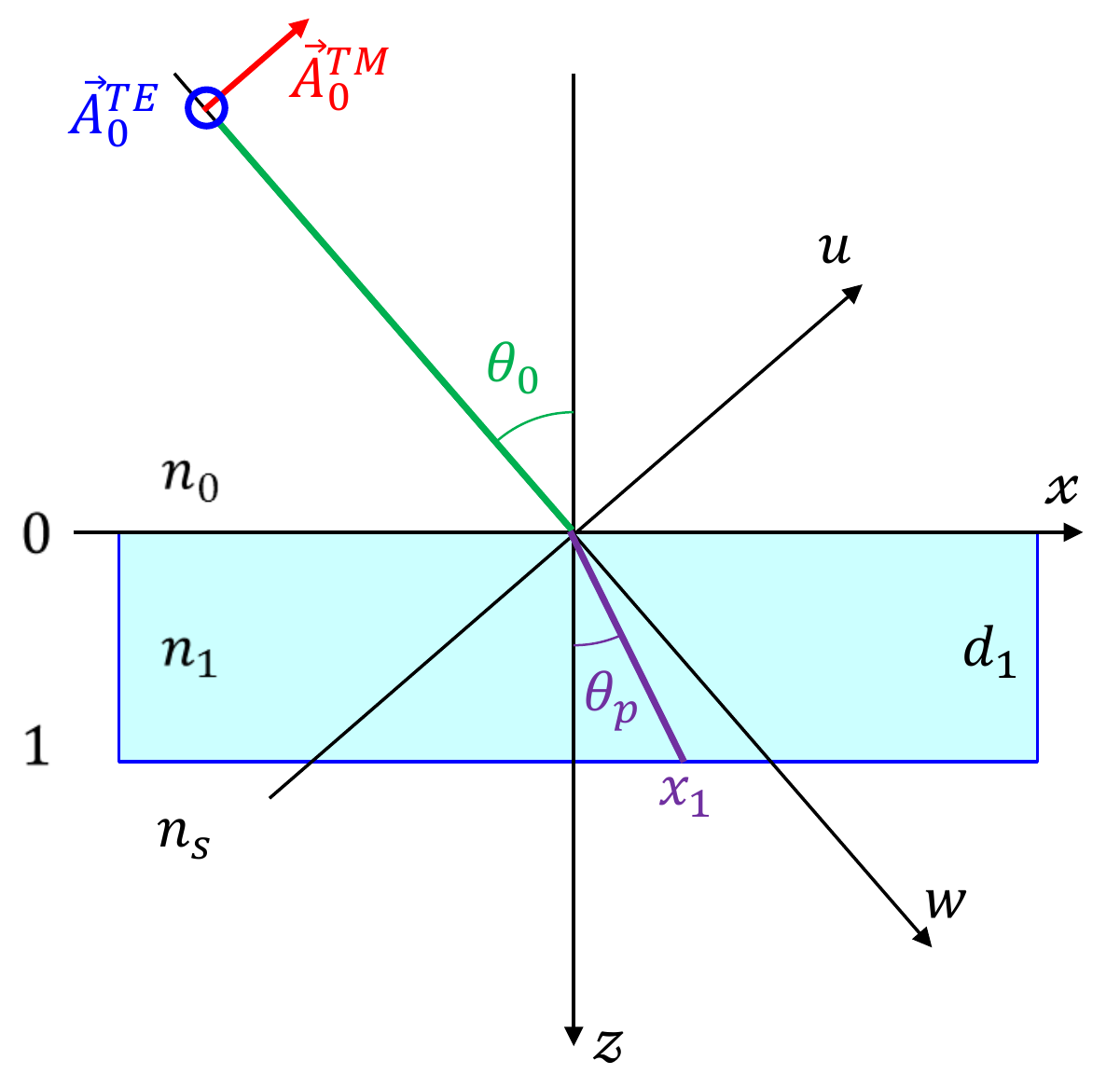}
\caption{Schematic representation of the problem ($n_s=n_0$ corresponds to a wafer and $n_s=n_1$ to a boundary between two semi-infinite media)}
\label{fig:Frames}
\end{figure}

If one intends to implement displacement measurement to achieve such a verification, limiting the lateral extension of the beam and, accordingly, not using a plane wave, but rather a wave-packet characterized by non null divergence, are absolutely required. Consider, for example, an incident Gaussian beam propagating along the $w$ axis; thus, the corresponding spatial field distribution in the $uvw$ frame is given by
\begin{equation}
\vec{\mathcal{E}}(u,v,w) =\iint\vec{\mathbb{A}}(\sigma_u,\sigma_v)\thinspace e^{i[\sigma_uu+\sigma_vv+\alpha_0(\sigma)w]}d\sigma_ud\sigma_v
\label{eq:Euvw}
\end{equation}
with
\begin{equation}
\vec{\mathbb{A}}(\sigma_u,\sigma_v)=\vec{\mathbb{A}}_0\thinspace e^{-\frac{w_0^2}{4}(\sigma_u^2+\sigma_v^2)}
\end{equation}
and
\begin{equation}
\alpha_0^2(\sigma)=k_0^2-(\sigma_u^2+\sigma_v^2)=k_0^2-\sigma^2
\label{eq:alpha}
\end{equation}
This Gaussian beam is impinging upon the upper face of a plane parallel wafer (index of refraction $n_1$, thickness $d_1$) lying in the $z=0$ plane. Moreover, we will assume without lack of generality that the origins of both frames $xyz$ and $uvw$ are coincident (see Fig. \ref{fig:Frames}). The angle of incidence (AOI) is thus defined by the angle $\theta_0$ between the O$z$ and O$w$ axes. For describing the propagation of the Gaussian beam in this new frame, we introduce a change of basis corresponding to a rotation of center O and angle $\theta_0$ in the $u$O$w$ plane, i.e.,
\begin{equation}
\left\{
\begin{aligned}
&u=x\cos\theta_0-z\sin\theta_0\\
&v=y\\
&w=x\sin\theta_0+z\cos\theta_0
\end{aligned}
\right.
\label{eq:BasisChange}
\end{equation}
By using (\ref{eq:BasisChange}), relation (\ref{eq:Euvw}) becomes
\begin{multline}
\vec{\mathcal{E}}(x,y,z) =\iint\vec{\mathbb{A}}(\sigma_u,\sigma_v)\thinspace e^{ix[\sigma_u\cos\theta_0+\alpha_0(\sigma)\sin\theta_0]}\\
e^{iy\sigma_v}e^{iz[\alpha_0(\sigma)\cos\theta_0-\sigma_u\sin\theta_0]\}}\thinspace d\sigma_u d\sigma_v
\label{eq:Exyz}
\end{multline}
which leads to the following distribution for the incident field in the $z=0$ plane
\begin{equation}
\vec{\mathcal{E}}_0^+(x,y) =\iint\vec{\mathbb{A}}(\sigma_u,\sigma_v)\thinspace e^{i\{x[\sigma_u\cos\theta_0+\alpha_0(\sigma)\sin\theta_0]+y\sigma_v\}}\thinspace d\sigma_u d\sigma_v
\label{eq:Exy0}
\end{equation}

To now describe the propagation of the Gaussian beam between the front face of the wafer ($z=0$) and its rear face ($z=d_1$), we must compute the complex admittance of the two interfaces \cite{Macleod2010Book,LequimeEDP2013}, i.e.,
\begin{equation}
Y_{1}=\tilde{n}_2=\tilde{n}_s\qquad Y_{0}=\frac{Y_1\cos\delta_1-i\tilde{n}_1\sin\delta_1}{\cos\delta_1-i\frac{Y_{1}}{\tilde{n}_1}\sin\delta_1}
\label{eq:AdmittancesComplexes}
\end{equation}
with
\begin{equation}
\hspace*{-0.7cm}
\tilde{n}_j=\frac{1}{\eta_0\mu_j}\left\{
\begin{aligned}
&n_j\alpha_j/k_j\quad \text{for TE polarization}\\
&n_jk_j/\alpha_j\quad \text{for TM polarization}
\end{aligned}
\right.\enskip j=0,1,2
\label{eq:EffectiveIndex}
\end{equation}
\begin{multline}
\alpha_j^2=k_j^2-(\sigma_x^2+\sigma_y^2)\\
=k_j^2-[\sigma_u\cos\theta_0+\alpha_0(\sigma)\sin\theta_0]^2-\sigma_v^2
\end{multline}
\begin{equation}
\hspace*{-6.7cm}
\delta_1=\alpha_1d_1
\end{equation}
In these equations, $\eta_0$ is the vacuum impedance ($\eta_0=\sqrt{\mu_0/\epsilon_0}$) and $\mu_j$ the relative magnetic permeability of the different media (here, $\mu_j=1$). The reflection $r$ and transmission $t$ are given by
\begin{equation}
r=\frac{\tilde{n}_0-Y_0}{\tilde{n}_0+Y_0}\qquad t = \frac{1+r}{\cos\delta_1-i\frac{Y_1}{\tilde{n}_1}\sin\delta_1}=|t|\thinspace e^{i\tau}
\label{eq:rt}
\end{equation}

The spatial distribution of the transmitted field in the $z=d_1$ plane is also simply given by
\begin{multline}
\vec{\mathcal{E}}_t^+(x,y)=\iint t(\theta_0,\sigma_u,\sigma_v)\vec{\mathbb{A}}(\sigma_u,\sigma_v)\\
\times e^{i\{x[\sigma_u\cos\theta_0+\alpha_0(\sigma)\sin\theta_0]+y\sigma_v\}}\thinspace d\sigma_u d\sigma_v
\label{eq:Exyt}
\end{multline}

Here we wish to only consider linear spatial dispersion effects, i.e. lateral beam shifts without deformation; for that, it is required to use low divergence light beams. Thus, we chose a beam waist ($w_0$) of approximately 300 $\mu$m, which corresponds, at wavelength $\lambda=650$ nm, to a divergence of approximately 0.7 mrad (2.5 arc minutes). Therefore, $\sigma_x$ and $\sigma_y$ are very small quantities and we can write, as a first approximation
\begin{equation}
\begin{aligned}
&|t|(\theta_0,\sigma_u,\sigma_v)\approx |t|(\theta_0,0,0)=|t(\theta_0)|\\
&\tau(\theta_0,\sigma_u,\sigma_v)\approx\tau(\theta_0)+\sigma_u\left.\frac{\partial\tau}{\partial\sigma_u}\right|_{\theta_0}+\sigma_v\left.\frac{\partial\tau}{\partial\sigma_v}\right|_{\theta_0}
\end{aligned}
\label{eq:ApproxTau}
\end{equation}
where the partial derivative with respect to $\sigma_v$ is null for symmetry reasons. Moreover, from (\ref{eq:alpha}), we have, at the same level of approximation
\begin{equation}
\alpha_0(\sigma)\approx k_0
\label{eq:ApproxAlpha0}
\end{equation}

By using (\ref{eq:ApproxTau}) and (\ref{eq:ApproxAlpha0}) in (\ref{eq:Exyt}), we obtain (the Fourier transform of a Gaussian is a Gaussian)
\begin{equation}
\vec{\mathcal{E}}_t^+(x,y)=\frac{4\pi}{w_0^2}\thinspace e^{ik_0x\sin\theta_0}t(\theta_0)\vec{\mathbb{A}}_0\thinspace e^{-\frac{[x\cos\theta_0+\frac{\partial\tau}{\partial\sigma_x}(\theta_0)]^2+y^2}{w_0^2}}
\end{equation}
The field distribution on the rear face of the wafer has a Gaussian shape, but with different waist sizes along the $x$ and $y$ axes ($w_0/\cos\theta_0$ and $w_0$, respectively). Moreover, the coordinates ($x_1,y_1$) of the beam center are as follows
\begin{equation}
x_1=-\frac{1}{\cos\theta_0}\frac{\partial\tau}{\partial\sigma_u}(\theta_0,0,0)=-\frac{1}{k_0\cos\theta_0}\frac{\partial\tau}{\partial\theta_0}\enskip;\enskip y_1=0
\label{eq:BeamCenterPosition}
\end{equation}

\subsection{Analytic refraction formula}
\label{sec:AnalyticFormula}

By referring to the sine law and the group index used in dispersive media, let us here introduce a “packet index” $n_p$, defined by (see Fig. \ref{fig:Frames})
\begin{equation}
n_p\sin\theta_p=n_0\sin\theta_0\quad\text{with}\quad x_1=d_1\tan\theta_p
\label{eq:PacketIndex}
\end{equation}
The analytical expression of the phase $\tau$ can be easily obtained by using (\ref{eq:AdmittancesComplexes}) and (\ref{eq:rt})
\begin{equation}
\tan\tau=\frac{\tilde{n}_1+\tilde{n}_0\tilde{n}_s/\tilde{n}_1}{\tilde{n}_0+\tilde{n}_s}\tan\delta_1
\label{eq:PhaseTau}
\end{equation}
which allows one to compute the expression of the packet index by combining (\ref{eq:BeamCenterPosition}), (\ref{eq:PhaseTau}), and (\ref{eq:PacketIndex}).

Consider first the case of a plane interface between two semi-infinite substrates ($n_s=n_1$). Thus, at the distance $d_1$ of the boundary
\begin{equation}
\tau=\delta_1=k_1d_1\cos\theta_1\enskip\Rightarrow\enskip\frac{\partial\tau}{\partial\theta_0}=-k_1d_1\sin\theta_1\frac{\partial\theta_1}{\partial\theta_0}
\end{equation}
and then, since $\partial\theta_1/\partial\theta_0=k_0\cos\theta_0/k_1\cos\theta_1$
\begin{equation}
x_1=-\frac{1}{k_0\cos\theta_0}\frac{\partial\tau}{\partial\theta_0}=d_1\tan\theta_1\quad\Rightarrow\quad n_p=n_1
\label{eq:ConsequenceSineLaw}
\end{equation}
Our theoretical approach leads in this case to the usual sine law, as expected. However, if we now consider the case of a plane and parallel window ($n_s=n_0\ne n_1$), the determination of the position of the beam center $x_1$ requires to differentiate relation (\ref{eq:PhaseTau}), which leads to
\begin{equation}
x_1=-\frac{\mathcal{B}}{k_0\cos\theta_0}\frac{\tan\delta_1}{1+\mathcal{A}^2\tan^2\delta_1}+d_1\tan\theta_1\frac{\mathcal{A}(1+\tan^2\delta_1)}{1+\mathcal{A}^2\tan^2\delta_1}
\label{eq:x1}
\end{equation}
with
\begin{equation}
 \mathcal{A}=\frac{1}{2}\left(\frac{\tilde{n}_1}{\tilde{n}_0}+\frac{\tilde{n}_0}{\tilde{n}_1}\right)
 \end{equation}
 and
 \begin{equation}
 \mathcal{B}=\frac{\partial\mathcal{A}}{\partial\theta_0}=\frac{\tilde{n}_0^2-\tilde{n}_1^2}{2\tilde{n}_0\tilde{n}_1}\left(\frac{1}{\tilde{n}_0}\frac{\partial\tilde{n}_0}{\partial\theta_0}-\frac{1}{\tilde{n}_1}\frac{\partial\tilde{n}_1}{\partial\theta_0}\right)
 \end{equation}

Such relation (\ref{eq:x1}) will be compared to that obtained in the case of a single boundary between two semi-infinite media and this will highlight the drastic influence of the interference phenomena induced by the partial reflection on the rear face of the wafer.
\section{Consequences}
\label{sec:Consequences}

\subsection{Free-standing thin film embedded in vacuum}
\label{sec:DielectricThinFilm}

Assume now that the thickness $d_1$ of the wafer is much smaller than the wavelength (for example, 25 nm), as considered by Dolling in \cite{Dolling2007OptExpr}. We can use the approximation $\tan\delta_1\approx\delta_1$, which allows us to compute the effective angle of refraction $\theta_p$
\begin{equation}
\tan\theta_p\approx\mathcal{A}\tan\theta_1-\mathcal{B}\frac{n_1\cos\theta_1}{n_0\cos\theta_0}
\end{equation}
Obviously, to finalize the calculation, we have to take into account the state of polarization of the incident light beam. For TE polarization, this leads to
\begin{equation}
\tan\theta_p^{\text{TE}}\approx\frac{1}{2}\tan\theta_0\left\{3-\frac{n_1^2-n_0^2\sin^2\theta_0}{n_0^2-n_0^2\sin^2\theta_0}\right\}
\end{equation}
whereas, for TM polarization, the effective angle of refraction $\theta_p^{\text{TM}}$ is given by
\begin{equation}
\tan\theta_p^{\text{TM}}\approx\frac{1}{2}\tan\theta_0\left\{\frac{2n_0^4+n_1^4}{n_0^2n_1^2}-\frac{n_1^2-n_0^2\sin^2\theta_0}{n_1^2-n_1^2\sin^2\theta_0}\right\}
\end{equation}
Thus, the interference induced by the reflection on the rear face gives rise to a double refraction phenomenon for an isotropic free-standing thin-film. Moreover, a negative refraction phenomenon can be observed for TE polarization if the angle of incidence $\theta_0$ satisfies the following condition
\begin{equation}
\theta_p^{\text{TE}}<0\enskip\Rightarrow\enskip\sin^2\theta_0 >\frac{3n_0^2-n_1^2}{2n_0^2}
\end{equation}
In other words, if the refractive index of the free-standing film is greater than $\sqrt{3}\thinspace n_0$, this negative refraction behavior can be observed at any angle of incidence.

In the same manner, for TM polarization
\begin{equation}
\theta_p^{\text{TM}}<0\enskip\Rightarrow\enskip\sin^2\theta_0>\sin^2\theta_c^{\text{TM}}=1-\frac{n_0^2(n_1^2-n_0^2)}{n_0^4+n_1^4}
\end{equation}
Thus, for angles of incidence greater than $\theta_c^{\text{TM}}$, negative refraction occurs simultaneously for both states of polarization.

\begin{figure}[htbp]
\centering
\includegraphics[width=.9\linewidth]{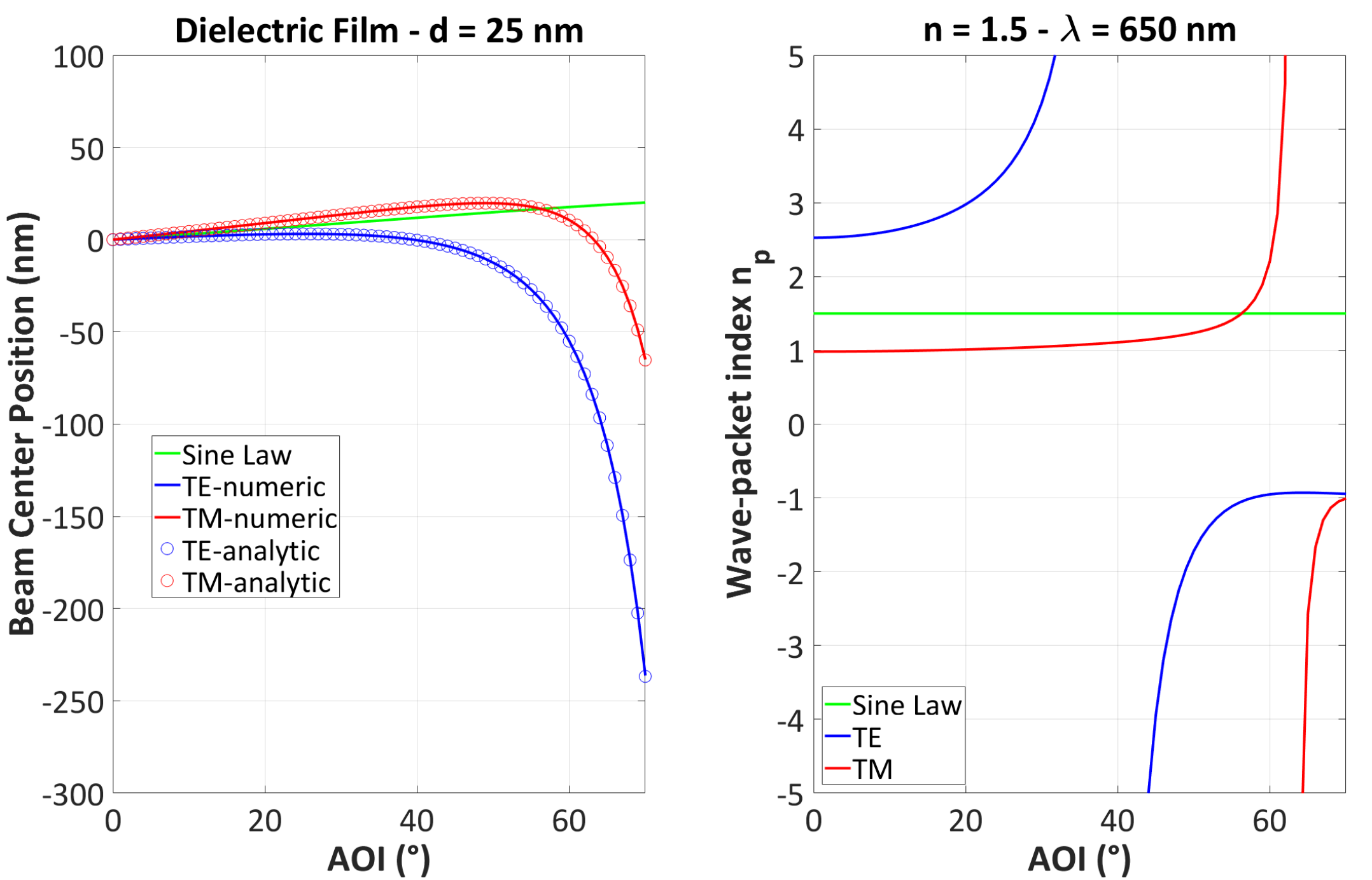}
\caption{Anomalous refraction behavior of a dielectric thin film ($d_1=25$ nm, $n_1=1.5$, $\lambda=650$ nm).\\
(Left graph, position of the beam center $x_1$ with respect to AOI; right graph, packet index $n_p$ with respect to AOI; blue curves, TE polarization; red curves, TM polarization; green curves, standard sine law)}
\label{fig:Dielectric_Thin_Film}
\end{figure}

Figure \ref{fig:Dielectric_Thin_Film} summarizes these results by showing the variation of the position of the beam center with respect to the angle of incidence for both states of polarization (left graph), by using either a numerical integration of the wave-packet given in (\ref{eq:Exyt}) or a direct application of the analytical relation (\ref{eq:x1}): one can note the quality of the accordance. The right graph of the same Figure shows the corresponding angular dependence of the packet indices $n_p^{\text{TE}}$ and $n_p^{\text{TM}}$.
\subsection{Flat lens regime}
\label{sec:FlatLensRegime}
In the flat lens regime \cite{Pendry2000PRL,Xu2013Nature}, the wave-packet index $n_p$ would be equal to -1 at any angle of incidence. To identify the conditions allowing that best satisfy this constraint, we chose to implement a dedicated discrepancy function $\text{DF}(n_1,d_1)$, defined by
\begin{equation}
\text{DF}(n_1,d_1)=\sqrt{\frac{1}{N}\sum\limits_{m=1}^N[n_p(\theta_m;n_1,d_1)+1]^2}
\end{equation}
where $n_p$ is computed with (\ref{eq:PacketIndex}) and (\ref{eq:x1}), while $\theta_m$ are regularly distributed AOIs over the range $[0,\theta_{\text{max}}=70$°], i.e. $\theta_m=\frac{m}{N}\theta_{\text{max}}$, $m=1,...,N$.
\begin{figure}[htbp]
\centering
\includegraphics[width=.8\linewidth]{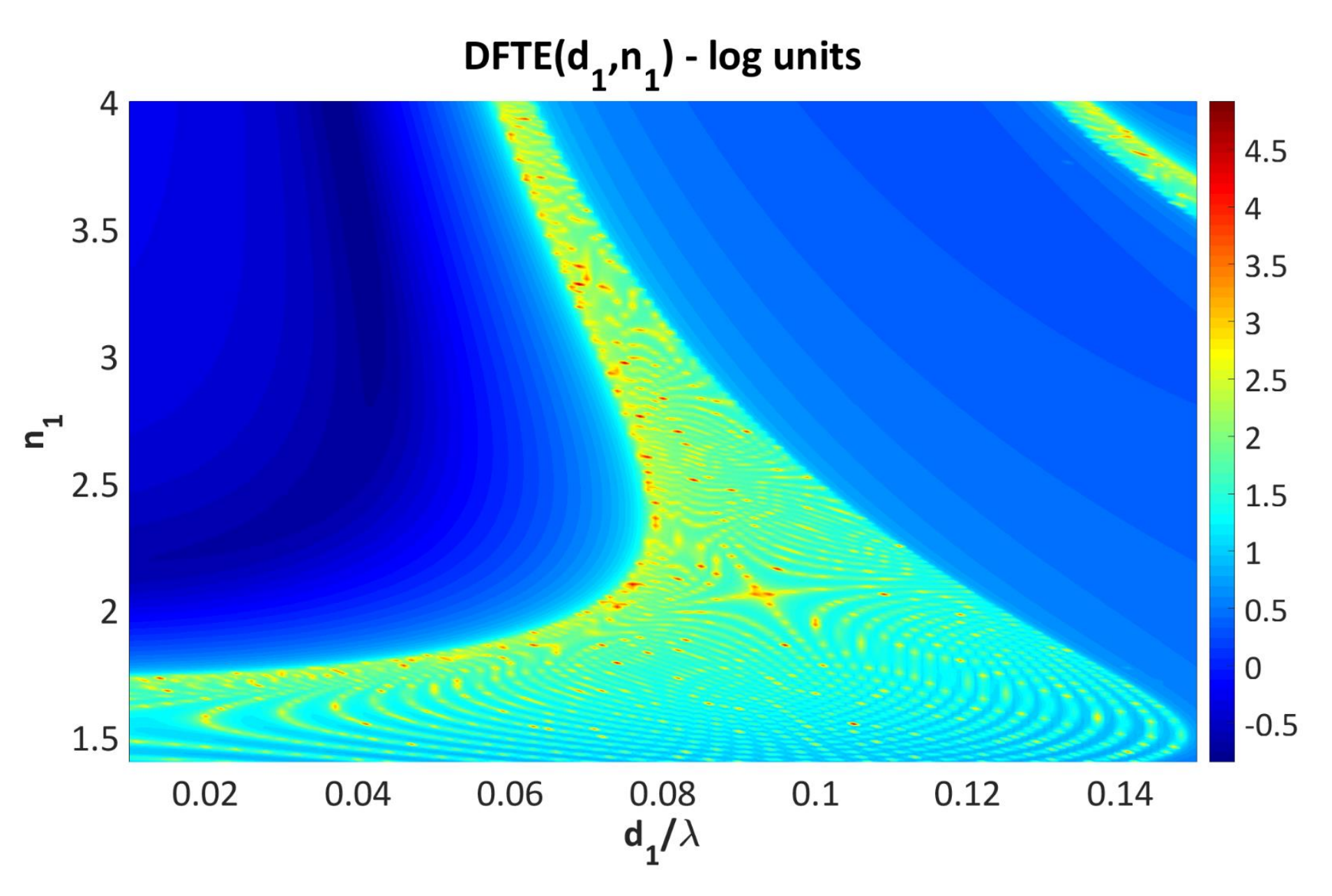}
\caption{Discrepancy function $\text{DF}(n_1,d_1)$ in TE polarization (false color representation in logarithmic units).}
\label{fig:Discrepancy_Function}
\end{figure}
Figure \ref{fig:Discrepancy_Function} shows that the best results are obtained for very thin films ($d_1<0.06\lambda$) and high refractive index ($n_1>2.5$). In practice, it is possible to be very close to an optimum configuration ($n_p^{\text{TE}} = -0.97\pm 0.1$) by using a 62-nm thick silicon film in its transparency region, at 1550 nm.

\begin{figure}[htbp]
\centering
\includegraphics[width=.9\linewidth]{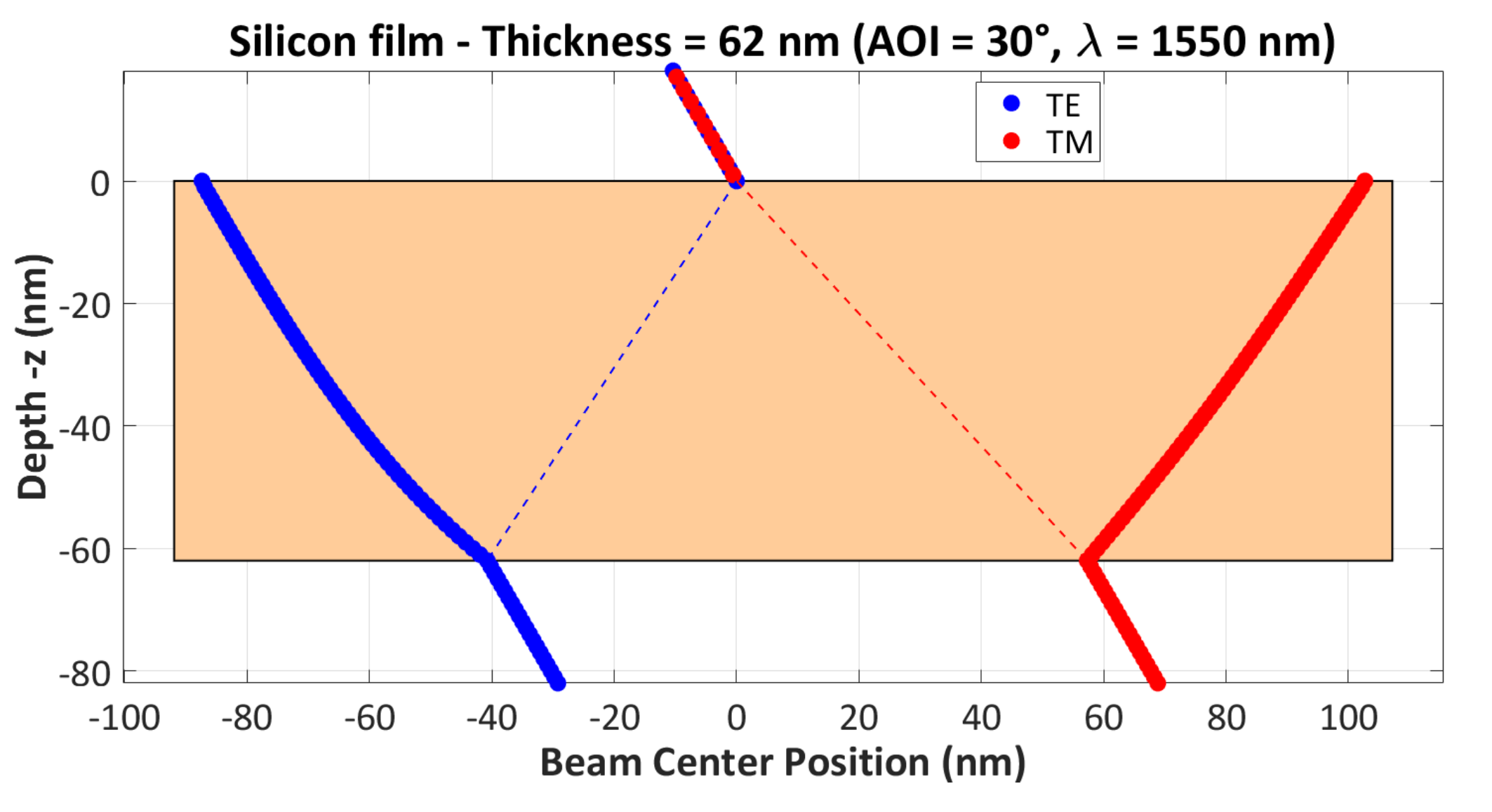}
\caption{Anomalous refraction behavior of a free-standing silicon thin film ($d_1=62$ nm, $n_1=3.48$, $\lambda=1550$ nm).\\
(Position of the beam center $x_1$ with respect to the depth $-z$ for a 30° AOI; blue dots, TE; red dots, TM)}
\label{fig:Silicon_Thin_Film}
\end{figure}

By replacing $d_1$ by $z$ in the approach described in Section \ref{sec:GeneralApproach}, one can also determine the position of the beam center at any depth inside the film. Figure \ref{fig:Silicon_Thin_Film} shows the result of this modeling for the silicon free-standing thin-film defined above. One can first note the large split induced between the two polarizations by the $(1+r)$ term at the top of the film. Dashed lines allow one to visualize the anomalous angle of refraction $\theta_p$, very close to $-\theta_0$ for TE polarization.
\subsection{Transparent window}
\label{sec:TransparentWindow}

When the thickness $d_1$ of the wafer increases, the second term of relation (\ref{eq:x1}) becomes predominant, and we have
\begin{equation}
x_1\approx \frac{\mathcal{A}(1+\tan^2\delta_1)}{1+\mathcal{A}^2\tan^2\delta_1}\thinspace d_1\tan\theta_1=\mathcal{K}(\delta_1)\thinspace d_1\tan\theta_1
\label{eq:x0Window}
\end{equation}
where $x_1=d_1\tan\theta_1$ corresponds to the standard sine law. The position of the beam center varies between $\mathcal{A}d_1\tan\theta_1$ and $(d_1/\mathcal{A})\tan\theta_1$ versus the thickness phase $\delta_1$ and the state of polarization of the incident light beam.

For TE polarization, $\mathcal{A}$ is an increasing function of the angle of incidence, and $\mathcal{A}=(n_0^2+n_1^2)/2n_0n_1>1$ at normal incidence; thus, $\mathcal{A}^{\text{TE}}>1$ for any AOI.
For TM polarization, $\mathcal{A}$ decreases first with AOI up to the Brewster angle before increasing. At Brewster incidence, $\mathcal{A}^{\text{TM}}=1$, and the effect of anomalous refraction disappears.

To visualize this effect, we search for a driving physical parameter adapted to the very large value of the wafer optical thickness: the wavelength of the light beam appears the best candidate. 
\begin{figure}[htbp]
\centering
\includegraphics[width=.95\linewidth]{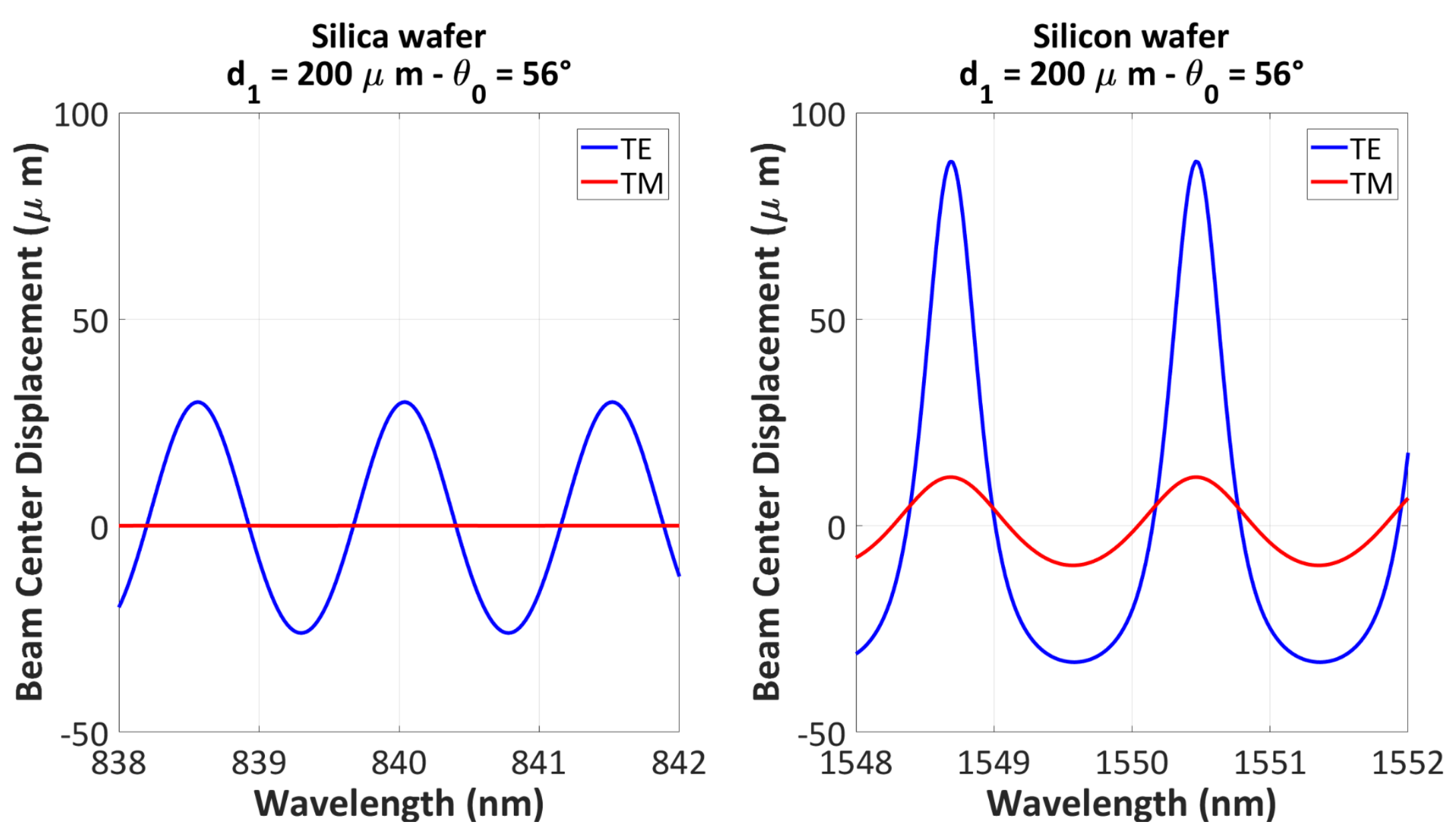}
\caption{Anomalous refraction behavior of 2 transparent wafers: beam center shift with respect to the position defined by the sine law (in $\mu$m) for a 56° AOI (Brewster angle for silica)\\
(Left graph, silica wafer at approximately 840 nm; right graph, silicon wafer at approximately 1550 nm; red solid lines, TE polarization; blue solid lines, TM polarization).}
\label{fig:Silica_Silicon_Wafer}
\end{figure}
Let us consider a 200-$\mu$m thick silica wafer illuminated at the Brewster angle (i.e. 56°) by a wavelength-tunable monochromatic light beam at approximately 840 nm. The left graph of Fig. \ref{fig:Silica_Silicon_Wafer} shows the evolution of the position of the beam center at the rear face of the wafer when changing this wavelength: these variations are illustrated through the shift between the actual position of the beam and the position defined by the classical sine law. As expected for a Brewster AOI, this shift is equal to zero for TM polarization. The right graph shows the same result, but for a 200 $\mu$m thick silicon wafer and a central wavelength of 1550 nm. As expected again, the increase of the refractive index induces a shape change of the periodic shift in TE polarization.
The differential values of the shifts are in both cases approximately a few tens of microns.

\section{Discussion}
\label{sec:Discussion}

One can be surprised that such noticeable effect (a few tens microns of relative shift) was not observed before; however, we must keep in mind that its visibility is canceled by a lack of flatness of the wafer or an excessively large spectral bandwidth of the light source. Indeed, by performing an integration of the relation (\ref{eq:x0Window}) over $2\pi$ variation of $\delta_1$, we find
\begin{equation}
\langle x_1\rangle_{2\pi}=\frac{1}{2\pi}\int\limits_{\delta_1-\pi}^{\delta_1+\pi}\mathcal{K}(\varphi)\thinspace d\varphi\times\thinspace d_1\tan\theta_1=d_1\tan\theta_1
\end{equation}

Thus, to achieve an experimental observation of this effect, one has to satisfy two main constraints. First, that the variation of the phase thickness $\delta_1$ over the beam footprint must be less that $\pi/10$, i.e., translated into wedge angle $\alpha$,
\begin{equation}
\frac{2\pi}{\lambda}n_1\cos\theta_1\frac{4w_0}{cos\theta_1}\thinspace\alpha\le\frac{\pi}{10}\quad\Rightarrow\quad\alpha\le 5''
\end{equation}
which is quite standard for wafer manufacturers. Second, the spectral bandwidth $\delta\lambda$ of the source must be specified by
\begin{equation}
2\pi\frac{\delta\lambda}{\lambda^2}n_1d_1\cos\theta_1\le\frac{\pi}{10}\quad\Rightarrow\quad\delta\lambda\le0.2\text{ nm}
\end{equation}
which is widely fulfilled with an external cavity semiconductor laser.

The experimental demonstration of this new effect can be achieved with the same type of experimental set-up as that implemented for the measurement of the Goos-Hänchen shift \cite{GoosAnnPhys1947}. This set-up is based on a differential measurement between the TE and TM shifts achieved through the modulation of the polarization state of a laser by an electro-optic modulator combined with a precise measurement of the resulting spatial displacement with a position-sensitive detector \cite{GillesOL2002,YallapragadaRSI2016}.

\newpage 

\renewcommand{\refname}{References with title and final page number}


\begin{thebibliography}{99}

\bibitem{Valentine2008Nature}
J. Valentine, S. Zhang, T. Zentgraf, E. Ulin-Avila, D. A. Genov, G. Bartal, and X. Zhang, “Three-dimensional optical metamaterial with a negative refractive index,” Nature \textbf{455}, 376-379 (2008).
\bibitem{Yu2011Science}
N. Yu, P. Genevet, M. A. Kats, F. Aieta, J.-P. Tetienne, F. Capasso, Z. Gaburro, “Light Propagation with Phase Discontinuities: Generalized Laws of Reflection and Refraction,” Science \textbf{334}, 333-337 (2011).
\bibitem{DollingThesis2007}
G. Dolling, “Design, Fabrication, and Characterization of Double-Negative Metamaterials for Photonics,” \textit{Dissertation zur Erlangung des Academischen Grades eines Doktors der Naturwissenschaften von des Fakultât für Physik der Universitât Karlsruhe} - https://publikationen.bibliothek.kit.edu/1000006964 (2007).
\bibitem{Dolling2007OptExpr}
G. Dolling, M. W. Klein, M. Wegener, A. Schädle, B. Kettner, S. Burger, and S. Linden, “Negative beam displacements from negative-index photonic metamaterials,” Opt. Express \textbf{15}, 14219-14227 (2007).
\bibitem{Macleod2010Book}
H. A. Macleod, \textit{Thin-Film Optical Filters}, 4th ed., CRC Press (2010).
\bibitem{LequimeEDP2013}
M. Lequime and C. Amra, \textit{De l'Optique Electromagnétique à l'Interférométrie - Concepts et Illustrations}, EDP Sciences (2013).
\bibitem{Pendry2000PRL}
J. B. Pendry, "Negative Refraction Makes a Perfect Lens," Phys. Rev. Lett. \textbf{85}, 3966-3969 (2000).
\bibitem{Xu2013Nature}
T. Xu, A. Agrawal, M. Abashin, K.J. Chau, and H. J. Lezec, “All-angle negative refraction and active flat lensing of ultraviolet light,” Nature \textbf{497}, 470-474 (2013).	
\bibitem{GoosAnnPhys1947}
F. Goos and H. Hänchen, “Ein neuer und fundamentaler Versuch zur Totalreflexion,” Ann. Phys. \textbf{436}, 333-346 (1947).
\bibitem{GillesOL2002}
H. Gilles, S. Girard, and J. Hamel, “Simple technique for measuring the Goos-Hänchen effect with polarization modulation and a position sensitive detector,” Opt. Lett. \textbf{27}, 1421-1423 (2002).	
\bibitem{YallapragadaRSI2016}
V. J. Yallapragada, G. L. Mulay, C.N. Rao, A. P. Ravishankar, and V. G. Achanta, “Direct measurement of the Goos-Hänchen shift using a scanning quadrant detector and a polarization maintaining fiber,” Rev. Sci. Instrum. \textbf{87}, 103109 (2016).

\end{thebibliography}
\end{document}